\documentclass[runningheads]{llncs}
\usepackage[T1]{fontenc}
\usepackage{graphicx,verbatim}

\usepackage[table,xcdraw]{xcolor}
\usepackage{amssymb}
\usepackage{graphicx} 
\usepackage{subcaption} 

\usepackage[colorlinks=true, urlcolor=blue, linkcolor=black]{hyperref}

\begin{document}
\title{A novel Fourier Adjacency Transformer for advanced EEG emotion recognition}

\titlerunning{Fourier Adjacency Transformer}

\author{Jinfeng Wang\inst{1}\thanks{These authors are contribute equally.}\orcidID{0000-0002-9086-7518}\and
Yanhao Huang\inst{2}\textsuperscript{$\star$}\and
Sifan Song\inst{2}\and
Boqian Wang\inst{2}\and
Jionglong Su\inst{2}\thanks{The corresponding authors.}\and
Jiaman Ding\inst{1}\textsuperscript{$\star\star$}
}
\authorrunning{J Wang et al.}
%
\institute{Faculty of Information Engineering and Automation, \\Kunming University of Science and Technology, Kunming, China \and
School of Artificial Intelligence and Advanced Computing, \\Xi’an Jiaotong-Liverpool University, Suzhou, China\\
\email{Jionglong.Su@xjtlu.edu.cn}\\
\email{Jiamanding@kust.edu.cn}}

\maketitle              

\begin{abstract}
EEG emotion recognition faces significant hurdles due to noise interference, signal nonstationarity, and the inherent complexity of brain activity which make accurately emotion classification. In this study, we present the Fourier Adjacency Transformer, a novel framework that seamlessly integrates Fourier-based periodic analysis with graph-driven structural modeling. Our method first leverages novel Fourier-inspired modules to extract periodic features from embedded EEG signals, effectively decoupling them from aperiodic components. Subsequently, we employ an adjacency attention scheme to reinforce universal inter-channel correlation patterns, coupling these patterns with their sample-based counterparts. Empirical evaluations on SEED and DEAP datasets demonstrate that our method surpasses existing state-of-the-art techniques, achieving an improvement of approximately 6.5\% in recognition accuracy. By unifying periodicity and structural insights, this framework offers a promising direction for future research in EEG emotion analysis. The code can be found at: \url{https://github.com/YanhaoHuang23/FAT}


\end{abstract}

\section{Introduction}
Electroencephalography (EEG) emotion recognition has gained significant attention in affective computing, brain-computer interfaces (BCIs), and mental health assessment due to its high temporal resolution and objective measurement of emotional states~\cite{alarcao2017emotions}. However, challenges such as non-stationarity, low signal-to-noise ratio, and inter-subject variability hinder robust and generalized emotion recognition~\cite{alarcao2017emotions,bashivan2015learning,zheng2017identifying,jenke2014feature}.

Existing approaches primarily rely on raw EEG signals, handcrafted features, or deep learning-based representations. While deep models operating on raw EEG data preserve the full temporal structure, they struggle with high dimensionality, noise sensitivity, and poor generalization~\cite{7104132,li2019domain}. Handcrafted feature-based methods, such as differential entropy (DE) and power spectral density (PSD), enhance interpretability and robustness but may lead to information loss~\cite{jenke2014feature,duan2013differential}. Despite the significance of periodic EEG features, their effective extraction and utilization in deep learning remain underexplored~\cite{liu2010real,li2016emotion}.

Deep learning-based methods can be categorized into temporal, spatial, and hybrid modeling approaches. Recurrent neural networks (RNNs), including LSTMs (Long Short-Term Memory) and GRUs (Gated Recurrent Units), capture temporal dependencies but suffer from vanishing gradients and limited long-range modeling~\cite{yang2018emotion,song2018eeg}. Convolutional neural networks (CNNs) extract localized spatial patterns but lack temporal modeling capacity~\cite{phan2021eeg}. Hybrid approaches, such as graph neural networks, leverage EEG’s topological structure but rely on predefined connectivity assumptions, limiting generalization across datasets~\cite{phan2021eeg,zhong2020eeg}.

Recently, Transformer-based models have demonstrated strong performance by leveraging self-attention to capture global dependencies~\cite{song2021transformer,feng2022eeg,song2022eeg}. However, their data inefficiency and lack of inductive bias hinder stability and generalization in EEG applications. Addressing these challenges, this study introduces a novel Transformer-based framework that enhances EEG feature extraction and improves generalization.

\begin{itemize}
    \item We propose Fourier Analytic Linear (FAL) Layer as a novel query-key-value mapping mechanism in self-attention, enabling the decoupling of periodic and aperiodic EEG representations to capture inherent periodic structures.

    \item By integrating FAL with multi-head self-attention (MHSA)~\cite{vaswani2017attention}, we propose Fourier Adjacent Attention (FAA) groups attention heads based on learned periodicity, significantly enhancing model performance.

    \item Replacing standard self-attention with FAA, we develop Fourier Adjacent Transformer (FAT), a novel EEG emotion recognition framework achieving state-of-the-art (SOTA) performance and superior generalization across SEED-family benchmarks and the DEAP dataset.

\end{itemize}

\section{Methods}
\subsection{Fourier Analytic Linear}

To effectively decouple and extract the periodic components in EEG signals, we propose a novel module, FAL, inspired by Fourier Analysis Networks (FAN)~\cite{dong2024fan}. FAN enhances the ability of neural networks to model periodic signals by incorporating a Fourier transform-based periodic modeling mechanism. However, within the FAN structure, the aperiodic component is typically processed through a nonlinear activation function, which may compromise its linear mapping capability, particularly in the Transformer architecture.

The key innovation of FAL lies in ensuring that the aperiodic component remains purely linear. This is achieved by replacing the nonlinear activation function applied to the aperiodic component in FAN with an Identity Function, thereby preserving its linearity. By maintaining a purely linear transformation for the aperiodic component, FAL can seamlessly replace the standard linear mapping layer in Transformer structures, ensuring compatibility with self-attention mechanisms while retaining the benefits of periodic signal modeling introduced by Fourier-Transform. The formulaic expression of FAL is

\begin{equation}
\label{equa: FAL}
    \Phi(x) = [cos(W_px)||sin(W_px)||W_nx+B_n],
\end{equation}
where $x \in \mathbb{R}^{d_{in}}$ and the $||$ refer to concatenate operation. The FAL consists of two main parts. First, the periodic part is obtained by unbiased linear transformation and Fourier basis function modeling:
\begin{equation}
\label{equa: P}
    \Phi_p(x) = [cos(W_px)||sin(W_px)].
\end{equation}

\begin{figure}[t]
    \centering
    \includegraphics[width=\textwidth]{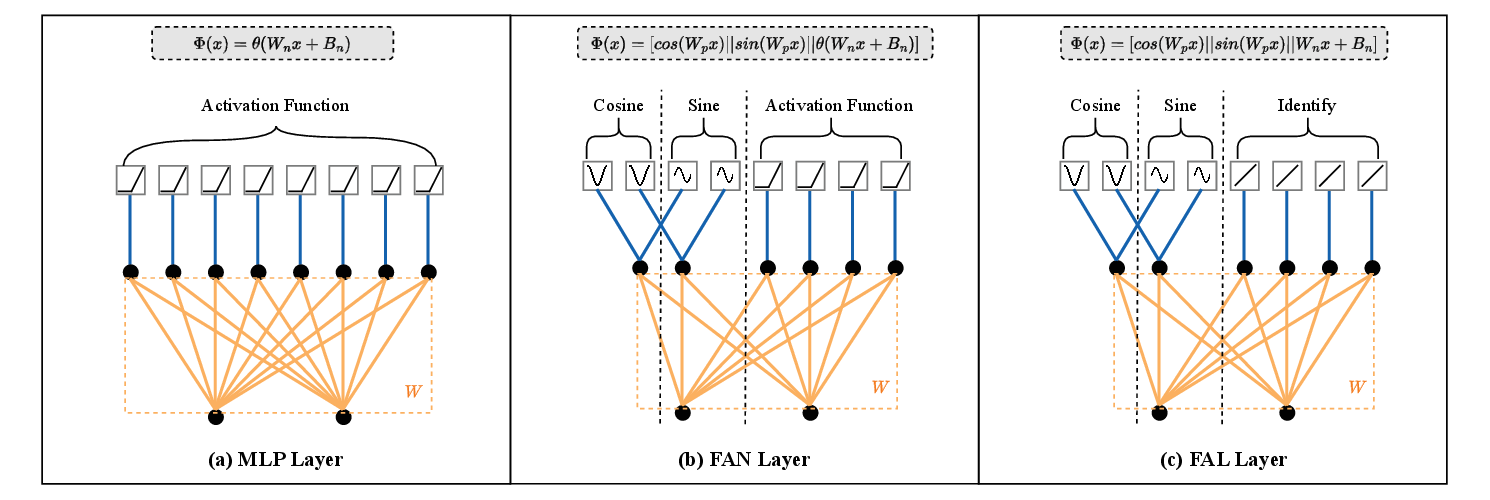}
    \caption{Illustrations of MLP Layer vs. FAN Layer vs. FAL Layer.}
    \label{fig:FAL}
\end{figure}

As shown in Fig.~\ref{fig:FAL}, in the aperiodic part, unlike FAN, we directly use a linear transformation thus ensuring its linear mapping capability, i.e.,
\begin{equation}
\label{equa: a}
    \Phi_a(x) = W_nx+B_n.
\end{equation}

This design enables FAL to function as a linear projection layer within the Transformer architecture. Its structure facilitates the decoupling of periodic and aperiodic representations along the channel dimension, providing greater flexibility for the subsequent MHSA. Generally, compared to the traditional MLP-based QKV (Query, Key and Value) projection, FAL offers a significant advantage by explicitly capturing the inherent periodic structure of EEG signals while maintaining high computational efficiency.

\subsection{Fourier Adjacent Attention}
\label{sec: FAA}

\begin{figure}[t]
    \centering
    \includegraphics[width=\linewidth]{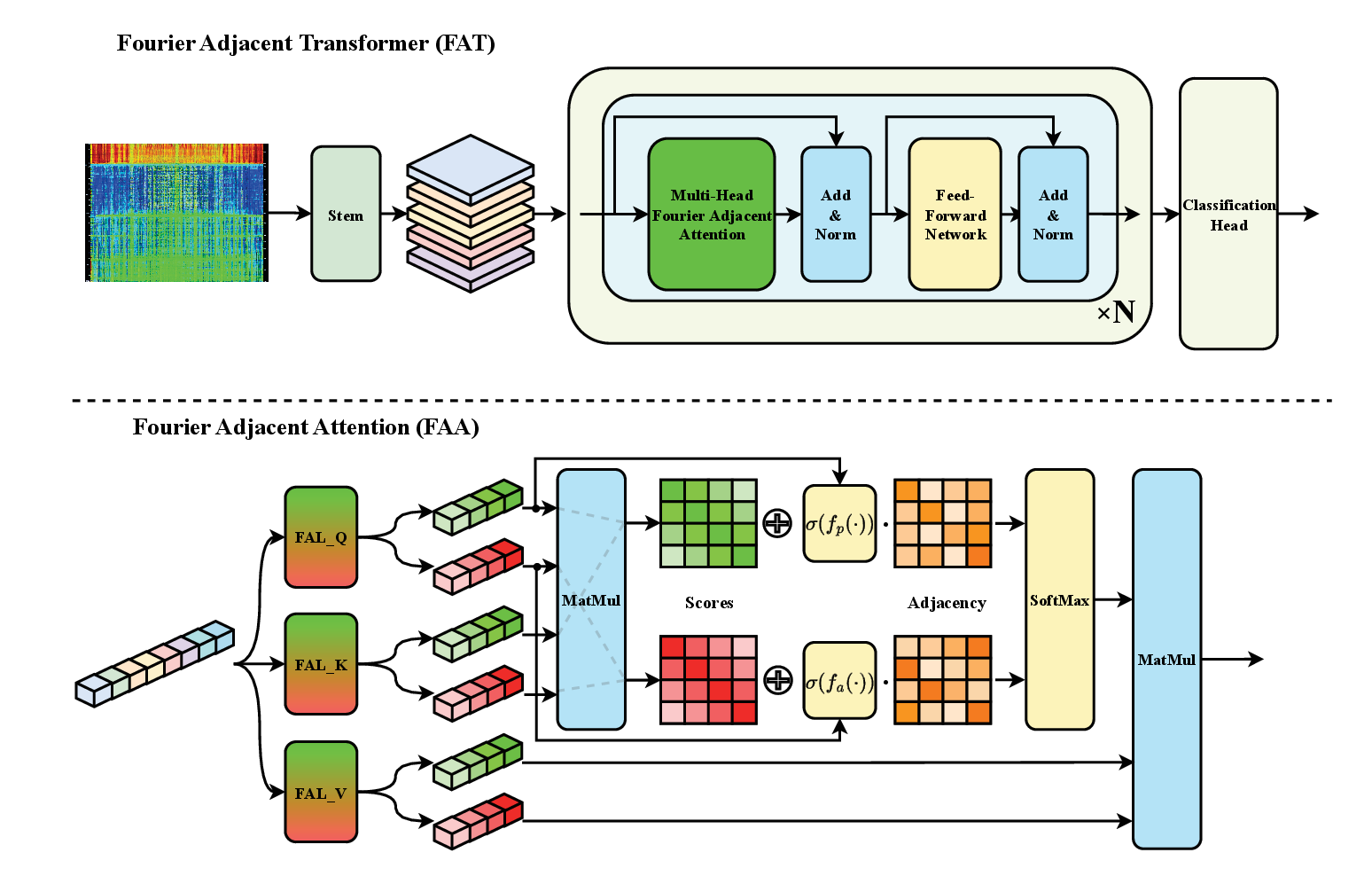}
    \caption{The overview of the FAT and FAA.}
    \label{fig:FAT}
\end{figure}

Furthermore, building upon FAL, we propose FAA to enhance the self-attention mechanism's ability to model periodic signals. FAA leverages FAL’s capability to disentangle periodic and aperiodic components while incorporating EEG inter-channel adjacencies into the attention computation process. This integration enables FAA to more effectively capture both universal and sample-specific connectivity patterns in EEG signals. Fig.~\ref{fig:FAT} illustrates the computational flow of FAA in the single-head case.

As shown in Fig.~\ref{fig:FAT}, in FAA, the FAL-based QKV projector transforms the input sequences into query, key, and value representations while simultaneously decoupling them into periodic and aperiodic components. Specifically, for each input feature vector $x$, the transformation results in

\begin{equation}
    Q = [Q_p || Q_a], \; K = [K_p || K_a], \; V = [V_p || V_a],
\end{equation}

\begin{equation}
    Q_p||Q_a = FAL\_Q(x), \; K_p||K_a = FAL\_K(x), \; V_p||V_a = FAL\_V(x),
\end{equation}
where $Q_p$ represents the periodic component of query, and $Q_a$ refer to the aperiodic component of  query. The same decomposition applies to Key and Value. This FAL-based projection ensures that FAA can independently model periodic and aperiodic information, providing a comprehensive representation for subsequent attention computations.

In FAA, since the query, key, and value representations are decoupled into periodic and aperiodic components, we apply scaled dot-product attention separately to each part, resulting in two distinct attention score matrices. Unlike traditional MHSA, which directly computes the attention matrix using the SoftMax function after obtaining the attention scores, FAA introduces a pair of learnable adjacency matrices with depth-sharing parameters. These matrices are then added to the attention score matrices through a gating module, and the resulting sum is fed into the SoftMax function to compute two final attention matrices $Attention_p$ and $Attention_a$ They can be formulated as follows:

\begin{equation}
    Attention_{p/a} = SoftMax(\frac{Q_{p/a}K_{p/a}^T}{\sqrt{d_k}} + \sigma(f_{p/a}(Q_{p/a}))\times Adjacency_{p/a}),
\end{equation}
where $f_{p/a}$ is implemented by a linear layer and $\sigma$ refers to the Sigmoid function. After deriving the two attention matrices $Attention_p$ and $Attention_a$, the final output of the FAA is calculated:
\begin{equation}
    O = [Attention_p\cdot V_p || Attention_a\cdot V_a].
\end{equation}

Another key contribution of FAA, beyond the incorporation of periodic representation processing, is the introduction of additive adjacency matrices in the computation of the attention matrix. We propose that this pair of adjacency matrices captures the structural correlations between EEG channels from periodic and aperiodic perspectives, respectively. By integrating them with the attention score matrices, which reflect data-driven, sample-specific correlations, FAA effectively mitigates the instability arising from the lack of inductive bias in traditional MHSA.

\subsection{Fourier Adjacent Transformer}
As shown in Fig.~\ref{fig:FAT}, we replace the self-attention mechanism of MHSA with FAA while retaining the multi-head design, forming a new module Multi-Head Fourier Adjacency Attention. We shall integrate this novel module into the Transformer architecture, leading to the proposed novel FAT model.

Since FAT retains the Transformer architecture, it requires sequencing and embedding of the raw EEG data. In this study, we use DE features of EEG as model inputs, followed by patch embedding and positional embedding to generate input representations suitable for multi-head FAA. The corresponding formula is expressed as follow:

\begin{equation}
    x = ReLU(BN(f_2(ReLU(BN(f_1(input))))))
\end{equation}
where $input\in \mathbb{R}^{B\times C\times F}$, $x\in \mathbb{R}^{B\times C\times E}$ and BN refers to batch normalization layer, $f_1$ ($F\to \frac{E}{2}$) and $f_2$ ($\frac{E}{2}\to E$) are represent linear layers. In this study, $F$ equals five, i.e., representing the five frequency bands of DE feature.

\subsection{Validation scheme}

To comprehensively evaluate the performance of FAT in the EEG emotion recognition task, we conduct validation through both subject-dependent and subject-independent experiments.

In the subject-dependent experiments, the training and test sets each contain a subset of data from all subjects in the dataset. This setup is designed to assess the learning and recognition capabilities of FAT. The data partitioning scheme strictly follows the configurations of various benchmark datasets.

In the subject-independent experiments, we adopt the Leave-One-Subject-Out (LOSO) approach, where the test set consists of data from subjects not included in the training set. This methodology is used to evaluate the generalization ability of FAT across unseen subjects.

\section{Results and discussions}

\subsection{Implementation details}

\subsubsection{Datasets}
In this study, to evaluate the performance of FAT across tasks of varying complexity, we selected SEED~\cite{zheng2015investigating,duan2013differential}, SEED-IV~\cite{8283814}, SEED-V~\cite{liu2021comparing}, and SEED-VII~\cite{10731546} from the SEED family, corresponding to three-class, four-class, five-class, and seven-class emotion classification tasks, respectively. Additionally, we utilized the widely recognized DEAP~\cite{koelstra2011deap} dataset, focusing on the classification of two key emotional states: Valence and Arousal.
In addition, all result data marked with * in the Table~\ref{tab:Subject-dependent} and Table~\ref{tab:Subject-independent} in this section are from the literature, and the bolded values are the best performance.
\subsubsection{Experimental settings}
Our method \textbf{is} implemented using PyTorch, and all experiments are conducted on NVIDIA A100 (80G) GPUs. For optimization, we employ Adam with a learning rate of 0.001 and a weight decay of 0.0001. The model is trained for 200 epochs, with a batch size of 64 for the SEED dataset and 32 for all other datasets. In the FAT, the number of attention heads (h) is set to eight, while the model depth is fixed at six.

For data augmentation, we randomly scale specific frequency bands of the frequency-domain features, restricting the scaling factor to (0.9, 1.1). To further enhance periodicity, we introduce a sine wave perturbation with a frequency of 2 and an amplitude of 0.05 in the frequency domain. Additionally, we employ Mixup, a data augmentation technique that linearly combines both samples and their corresponding labels, to further improve the model’s generalization ability.

\subsection{Recognition capability}

The experimental results in Table~\ref{tab:Subject-dependent} demonstrate that the proposed FAT consistently outperforms existing SOTA methods across the SEED family and DEAP datasets. Specifically, FAT achieves performance improvements of 2.9\%, 6.5\%, 6.6\%, 5.4\%, and 6.2\% over the previous SOTA on SEED-IV, SEED-V, SEED-VII, DEAP-Valence, and DEAP-Arousal, respectively, while maintaining competitive performance on the SEED dataset. These results underscore the effectiveness of FAT in learning discriminative representations and its strong capability in addressing complex recognition tasks.

\begin{table}[h]
    \centering
    \caption{Subject-dependent performance comparison. }
    \label{tab:Subject-dependent}
    \resizebox{\textwidth}{!}{ 
        \begin{tabular}{c|c|c|c|c|c|c}
            \hline
            Method    & SEED               & SEED IV           & SEED V           & SEED VII         & DEAP-Valence     & DEAP-Arousal     \\ \hline
            Scheme    & 9:6               & 16:8             & 3-fold           & 4-fold           & 10-fold          & 10-fold          \\ \hline
            KNN~\cite{cover1967nearest}       & 80.88/9.73         & 55.82/18.82       & 61.03/12.24      & 36.87/4.41       & 67.40/7.28               & 67.80/6.94               \\ \hline
            SVM~\cite{suykens1999least}       & 83.99/8.79         & 58.57/20.34       & 67.38/11.72      & 40.07/8.61       & 65.80/4.62        & 69.89/5.60        \\ \hline
            LSTM~\cite{graves2012long}      & --                 & --                & --               & --               & 74.76/3.41       & 71.81/3.99       \\ \hline
            DGCNN~\cite{song2018eeg}     & 89.75/8.51         & 69.09/13.79       & 69.52/9.57       & 46.36/7.33       & 84.69/6.20  & 83.02/7.07 \\ \hline
            BiHDM~\cite{li2020novel}     & 93.07/8.39         & 73.38/15.75       & 63.36/10.27      & 45.57/8.75       & --               & --               \\ \hline
            RGNN~\cite{zhong2020eeg,chen2024ugan,10731546}      & 94.24/5.95*   & 79.37/10.54* & 54.72/14.30       & 48.5/6.83*        & --               & --               \\ \hline
            PGCN~\cite{jin2024pgcn}      & \textbf{94.30/6.19} & 78.63/17.19       & 73.52/10.60       & 49.94/7.48       & --               & --               \\ \hline
            Conformer~\cite{song2022eeg} & 88.68/9.97         & 76.7/16.57        & 78.09/9.3  & 54.72/6.98 & 80.71/2.30        & 79.42/3.03       \\ \hline
            FAT (w/ FAN) & 93.18/11.28 & 79.03/16.27 & 82.95/8.64 & 59.42/8.59 & 80.47/2.89 & 83.09/3.85 \\ \hline
            FAT (ours) & 92.10/6.46 & \textbf{82.30/13.21} & \textbf{84.57/8.3} & \textbf{61.30/9.81} & \textbf{90.10/2.71} & \textbf{89.18/3.50} \\ \hline
        \end{tabular}
    }
\end{table}

\subsection{Generalization capability}

As shown in Table~\ref{tab:Subject-independent}, our proposed FAT demonstrates exceptional generalization capability compared to the current SOTA methods. Notably, on the most complex SEED-VII dataset, FAT outperforms Conformer, a leading SOTA method based on the same Transformer architecture, by approximately 5\%. We believe this consistent improvement suggests that FAT effectively learns to distinguish between universal and sample-specific information in EEG signals.

\begin{table}[h]
    \centering
    \caption{Subject-independent Performance comparison.}
    \label{tab:Subject-independent}
    \resizebox{\textwidth}{!}{ 
        \begin{tabular}{c|c|c|c|c|c|c}
            \hline
            Method    & SEED             & SEED IV          & SEED V           & SEED VII         & DEAP-Valence     & DEAP-Arousal     \\ \hline
            KNN~\cite{cover1967nearest}       & 48.4/12.50        & 32.92/8.08       & 30.26/16.67      & 21.01/4.67       & 55.35/4.3               & 52.86/4.22               \\ \hline
            SVM~\cite{suykens1999least}       & 56.48/15.20       & 36.28/12.99      & 39.49/16.20       & 30.73/11.12      & 54.7/0.36        & 55.9/0.76        \\ \hline
            LSTM~\cite{graves2012long}      & --               & --               & --               & --               & 60.31/0.24       & 59.78/0.27       \\ \hline
            DGCNN~\cite{song2018eeg}     & 77.29/9.71       & 53.84/9.87       & 64.05/15.84      & 35.94/7.88       & 59.72/1.38       & 57.26/1.29       \\ \hline
            BiHDM~\cite{li2020novel}     & 84.17/8.74       & 68.7/13.42       & 57.83/17.44      & 34.34/6.18       & --               & --               \\ \hline
            RGNN~\cite{zhong2020eeg,chen2024ugan,10731546}      & 85.30/6.72*  & 73.84/8.02*  & 66.28/16.71*      & 37.49/5.44*       & --               & --               \\ \hline
            PGCN~\cite{jin2024pgcn}      & 83.58/7.36 & 70.22/11.46     & 63.38/8.88       & 38.78/5.39       & --               & --               \\ \hline
            Conformer~\cite{song2022eeg} & 79.16/7.68       & 67.07/11.22      & 59.95/13.50       & 45.60/9.74        & 62.4/0.20         & 59.9/0.6         \\ \hline
            FAT (w/ FAN) & 83.24/5.58 & 70.52/11.69 & 64.66/14.49 & 49/10.32 & 62.59/0.64 & 60.00/0.48 \\ \hline
            FAT (ours) & \textbf{85.68/4.16} & \textbf{73.94/10.42} & \textbf{71.22/16.21} & \textbf{51.12/8.51} & \textbf{62.70/0.80} & \textbf{60.00/0.40} \\ \hline
        \end{tabular}
    }
\end{table}

\subsection{Ablation study}
To determine the optimal configuration of FAT, we conduct comprehensive ablation experiments on both data and model configurations.

The experimental results in Table~\ref{tab:performance_bands} show that the feature richness of the input data is positively correlated with the performance of the model. 

\begin{table}
    \centering
    
    \begin{minipage}{0.48\linewidth} 
        \centering
        \caption{Performance comparison for different frequency bands.}
        \label{tab:performance_bands}
        \resizebox{\textwidth}{!}{
            \begin{tabular}{c|c|c|c}
            \hline
            Band                     & SEED IV     & SEED V      & SEED VII    \\ \hline
            $\delta$ band            & 59.79/9.73  & 51.66/11.37 & 40.97/7.77  \\ \hline
            $\theta$ band            & 57.08/11.59 & 55.58/11.41 & 41.34/5.45  \\ \hline
            $\alpha$ band            & 53.3/10.13  & 58.69/12.36 & 38.53/6.31  \\ \hline
            $\beta$ band             & 55.57/10.24 & 58.58/14.92 & 42.31/5.77  \\ \hline
            $\gamma$ band            & 51.89/9.61  & 54.29/13.11 & 41.03/8.65  \\ \hline
            $\beta,\gamma$ bands     & 59.12/10.74   & 59.14/14.65 & 46.9/10.7 \\ \hline
            $\beta,\gamma,\delta$ bands & 67.5/11.43  & 61.77/13.65 & 48.16/10.19 \\ \hline
            All bands                & 73.94/10.42 & 71.22/16.21 & 51.12/8.51  \\ \hline
        \end{tabular}
        }
    \end{minipage}
    \hfill
    \begin{minipage}{0.48\linewidth}
        \centering
        \caption{Ablation studies on configurations and hyperparameters.}
        \label{tab:ablation}
        \resizebox{\textwidth}{!}{ 
            \begin{tabular}{c|cccc|c}
            \hline
            AdA & \multicolumn{4}{c|}{P-ratio} & LOSO Performance \\ \hline
               & 0 & 0.125 & 0.25 & 0.375 &  \\ \hline
               & \checkmark &  &  &  & 67.78/9.69  \\ \hline
            \checkmark & \checkmark &  &  &  & 68.97/11.1  \\ \hline
            \checkmark &  & \checkmark &  &  & 69.33/11.95 \\ \hline
            \checkmark &  &  & \checkmark &  & \textbf{73.94/10.42} \\ \hline
                       &  &  & \checkmark &  & 70.30/9.89 \\ \hline
            \checkmark &  &  &  & \checkmark & 69.03/10.96 \\ \hline
        \end{tabular}
        }
    \end{minipage}
\end{table}

In Table~\ref{tab:ablation}, AdA refers to the additive adjacency matrix. Following the FAN configuration, P-ratio represents the ratio of the output dimensions of the cosine and sine basis functions in FAL to the total output dimensions of FAL. Given that this study employs eight attention heads and an embedding dimension of 256, the possible values for P-ratio are 0, 0.125, 0.25, and 0.375. When P-ratio is set to 0, FAL degenerates into a linear layer. Experimental results indicate that FAL significantly enhances the performance of FAT. Furthermore, the periodic feature extraction capability of FAL helps to stimulate the additivity adjacency matrix, enabling the model to achieve superior accuracy.

\subsection{Visualization}
\begin{figure}[t]
    \centering
    \begin{subfigure}[b]{0.45\linewidth}
        \centering
        \includegraphics[width=\linewidth]{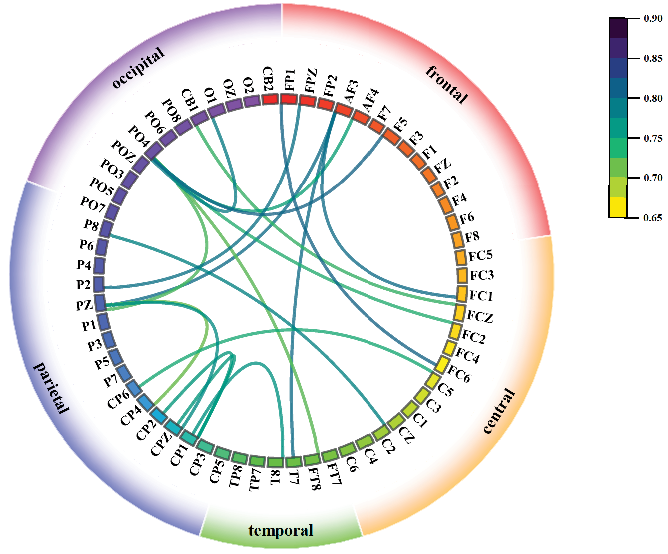}
        \caption{Periodic part}
        \label{fig:AdA_p}
    \end{subfigure}
    \hfill
    \begin{subfigure}[b]{0.45\linewidth}
        \centering
        \includegraphics[width=\linewidth]{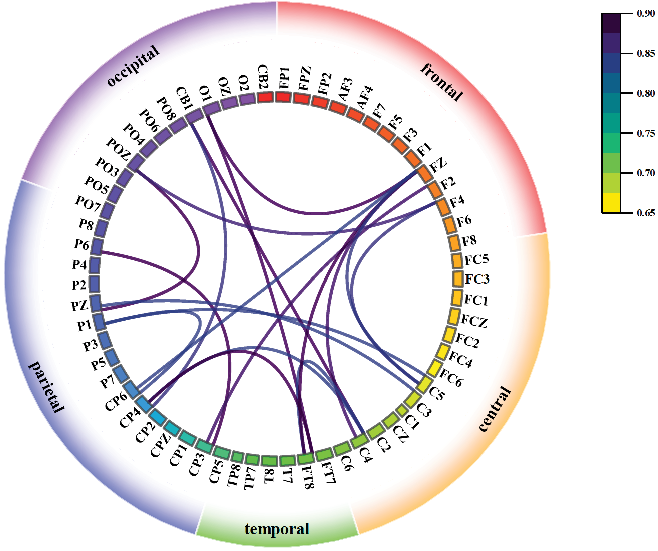}
        \caption{Aperiodic part}
        \label{fig:AdA_a}
    \end{subfigure}
    \caption{The top 15 correlations in each of the two adjacency matrices.}
    \label{fig:AdA}
\end{figure}

The adjacency matrix we introduced enables FAT to effectively capture EEG inter-channel relationships, similar to GNNs, thereby enhancing the model's interpretability. As shown in Fig.~\ref{fig:AdA}, the adjacency matrices of both the periodic and aperiodic components reveal distinct inter-channel structural relationships. Notably, the structural patterns learned from these two components exhibit significant differences, highlighting the model’s ability to extract diverse features from different signal properties.

\section{Conclusion}
In this paper, we propose a novel EEG emotion recognition model, the Fourier Adjacency Transformer, whose core component is Fourier Adjacency Attention. In Neighbor Attention, we extract and decouple the periodic and aperiodic representations in EEG signals while explicitly learning both universal and sample-specific connection patterns between EEG channels. This is achieved by introducing a adjacency matrix into the attention computation process. Experimental results demonstrate that FAT consistently outperforms other SOTA methods in terms of learning capability, recognition accuracy, and generalization. These findings highlight the potential of incorporating Fourier-based cognition into EEG analysis, and we plan to further explore this direction in future.

%
%
%
%

\end{document}